\documentclass{PoS}

\title{Multi-Objective Genetic Algorithm Optimisation for an Array of Cherenkov Telescopes}

\ShortTitle{Optimisation of an Array of Cherenkov Telescopes}

\author{Bruno Fontes Souto\thanks{Corresponding author.}\\
        Centro Brasileiro de Pesquisas F\'{i}sicas, 22290-180 URCA, Rio de Janeiro (RJ), Brazil\\
        E-mail: \email{bsouto@cbpf.br}}
\author{Ulisses Barres de Almeida\\
        Centro Brasileiro de Pesquisas F\'{i}sicas, 22290-180 URCA, Rio de Janeiro (RJ), Brazil\\
        E-mail: \email{ulisses@cbpf.br}}        
\author{\speaker{Ugo Giaccari}\\
        Centro Brasileiro de Pesquisas F\'{i}sicas, 22290-180 URCA, Rio de Janeiro (RJ), Brazil\\
        E-mail: \email{ugo@if.ufrj.br}}

\abstract{In the context of the development of the Cherenkov Telescope Array, we have conceived and implemented a multi-objective genetic algorithm (GA) code for the optimisation of an array of Imaging Atmospheric Cherenkov Telescopes (IACTs). The algorithm takes as input a series of cost functions (metrics) each describing a different objetive of the optimisation (such as effective area, angular resolution, etc.), all of which are expressed in terms of the relative position of the telescopes in the plane. The output of the algorithm is a family of geometrical arrangements which correspond to the complete set of solutions to the array optimisation problem, and differ from each other according to the relative weight given to each of the (maybe conflicting) objetives of the optimisation. Since the algorithm works with parallel optimisation it admits as many cost functions as desired, and can incorporate constraints such as budget (cost cap) for the array and topological limitations of the terrain, like geographical accidents where telescopes cannot be installed. It also admits different types of telescopes (hybrid arrays) and the number of telescopes of each type can be treated as a parameter to be optimised - constrained, for example, by the cost of each type or the energy range of interest. The purpose of the algorithm, which converges fast to optimised solutions (if compared to the time for a complete Monte Carlo Simulation of a single configuration), is to provide a tool to investigate the full parameter space of possible geometries, and help in designing complex arrays. It does not substitute a detailed Monte Carlo study, but aims to guide it. In the examples of arrays shown here we have used as metrics simple heuristic expressions describing the fundamentals of the IAC technique, but these input functions can be made as detailed or complex as desired for a given experiment. It is important to stress that the individual characteristics of each telescope are taken as fixed, and only the telescope arrangement is being optimised. Preliminary results will be presented in this contribution for the first time.}

\FullConference{35th International Cosmic Ray Conference -ICRC217-\\
		10-20 July, 2017\\
		Bexco, Busan, Korea}

\begin{document}

\section{Introduction}
The design and optimisation of large arrays of Cherenkov Telescopes has been extensively discussed in the context of the preparations for the Cherenkov Telescope Array (CTA)~\cite{ref1.1}. Given the complexity of modern instruments and the large parametric space to be considered, current studies greatly surpass the depth and extent of considerations taken into account for the design of previous generations of instruments~\cite{ref1.2}, such as the concept of juxtaposition of optimised cells of Cherenkov telescopes to compose a larger array~\cite{ref1.3, ref1.4}. Complete Monte Carlo studies~\cite{ref1.5} are the most adequate tool for investigating and designing best array layout candidates, but are computationally expensive and time-consuming. As a consequence, only part of the parametric space involved in the problem, can be effectively searched for with detailed simulations. 

Heuristic optimisation tools and software which allow for an approximate, but fast and ample exploration of the parametric space in question can be a valuable tool in guiding the design of large arrays such as CTA or other experiments in astroparticle physics. Algorithms such as Genetic Programming allow for multi-parametric optimisation of conflicting objectives and can handle, within a same framework, constraints external to the physics of the problem, but relevant to the search of an optimised solution, such as limitations in the topology of the terrain or cost restrictions.

Here we will present a preliminary implementation of such an algorithm designed to search for optimal layout configurations of an array of Cherenkov Telescopes. The work is conducted over a simplified toy model array, specifically built for this application, and does not represent any real instrument's implementation proposal; it serves as an example of how the algorithm can be used to inspect, in an efficient and complete way, the parametric space of the optimisation problem. The aim of the tool we developed is to present the array designer with a global perspective of potential solutions, intended to serve as a guide prior to more in depth investigation.

In this work we treated telescope properties as fixed a priori, not interfering with any instrumental parameters such as pixel size, telescope collection area, or trigger and analysis strategies for gamma-hadron separation. All optimisation regards simple considerations about the relative arrangement of instruments on the ground. Although the Evolutionary Algorithm is concluded, a more detailed and complete array to model is being currently developed.

In the following section we will present our toy model array and the metrics which will feed the physics of the problem into the optimisation algorithm. A brief description of Evolutionary Algorithms will be presented in Section 3, and the final section of the will show some simple applications and results to exemplify the work and its potential.

\section{Toy Model Array}

The fundamental assumption of this work is that the relevant physics for the design of a toy model array of Cherenkov telescopes can be effectively encapsulated in a set of cost functions. Together they provide a working model which to input in the optimisation algorithm. 

A final set of metrics is still under investigation, but since our primary purpose here is to present the optimisation algorithm itself, a simplified, preliminary toy model is considered, instead of a more detailed physical description. Care is taken, nevertheless, to ensure that the basic physics of the problem is described in a consistent way, so that results can be properly interpreted and the potential of the method correctly grasped. It is not our purpose to provide results that can be directly compared to real array design configurations. For this reason, no modelled array response functions or performance curves are presented, but only our algorithm's exploration of the parametric space for the array layout configurations will be shown 

Once the characteristics of the individual telescopes are held fixed, the problem of optimisation is reduced to that of the array layout itself. Provided that the instruments work in stereoscopic mode, the physical scale governing their relative arrangement on the ground is given by the size of the air shower's Cherenkov light pool, and the density and lateral distribution function of the Cherenkov photons. These in turn are functions of the primary gamma-ray energy and zenith angle of incidence, and subject to stochastic fluctuations of the shower development~\cite{ref2.1, ref2.2}. 

The metrics describing the toy model performance should be written in terms of the shower's impact parameter from a given telescope's position on the ground. This translates, in turn, to the relative spacing and arrangement between telescopes when we consider performance of the array stereoscopic detection. The optimisation procedure depends on a choice of the minimum number of telescopes required for an acceptable stereoscopic reconstruction of the shower events, which we will adopt to be equal to 3 in the examples shown here. 

For this demonstration, two \emph{heuristic expressions} were derived, which describe, in the simplest possible way, the objectives of the optimisation problem. For some description of the physics of the problem see~\cite{ref2.3}. Our metrics will be as as follows:\\

\noindent (i) The \emph{Effective Area} expresses the probability of detection of events by the array, and its maximisation is proportional to maximising the array photon collection and, consequently, sensitivity. It is given by the integral over the entire array area of the product of the geometric area element $dA$ and a probability function $f$. The latter indicates the probability of detection of an event with impact point at an area element $dA_i$ by a minimum number of $n$ telescopes of the array, where $n$ is the minimum stereoscopic multiplicity required for event reconstruction. So we can write,

\begin{equation}
A_{eff,~n} = \int_{i} f_i^{(n)}~dA
\end{equation}

\noindent (ii) The \emph{Angular Resolution} is associated to the quality of reconstruction of the shower direction and depends, mainly, on the shower viewing angle and the stereo angle. The former is a function of zenith angle (but for the moment we are neglecting this dependency, considering only near-vertical showers) and the impact parameter. 

Maximising the stereo angle $\theta_{i,j}$, refers to the condition of a maximum angular separation between the lines of sights of two telescopes $t_i$ and $t_j$ viewing a same shower event. This condition improves stereo reconstruction and is equivalent to having the telescopes isotropically distributed around the event. It corresponds to selecting the so-called "sweet spot" for stereo reconstruction: events falling within the region internal to the perimeter delimited by the telescopes. Events falling outside this "sweet spot" are worse reconstructed, even if viewed by a same number of instruments. 

In principle, each pair of viewing angles could be weighted by the Hillas parameters (size $s$, length $l$ and width $w$) of the shower image ellipse, in order to take into consideration the image quality for each instrument, but this additional step is beyond the considerations of this work. We adopt therefore the following expression:

\begin{equation}
\Gamma = sin \left[  \frac{ \sum \theta_{i,j}   }{N}   \right], \: \theta_{i,j}  \in  [0^{\circ},90^{\circ}],
\end{equation}

This physical model still lacks an adequate description for the shower image reconstruction (Hillas parameters), which impact not only on the shower direction reconstruction above, but also the energy reconstruction and background discrimination, as well as on the optimisation of the energy threshold. It is nevertheless sufficient for a self-consistent discussion of the Algorithm.

\section{Multi-objective Evolutionary Optimisation}

In the previous section we derived a pair of metrics to characterise a simplified toy model for an array of Cherenkov telescopes. All expressions were written in terms of the relative position between the telescopes on the ground. This group of functions will now serve as independent cost functions, each describing an objetive we want to optimise in planning the array. 

The major strength of the Evolutionary Algorithm described here is that it is able to handle all objectives in parallel and generate a family (or population) of solutions which cover the full spectrum of optimised telescope arrangements. This population of layout solutions, form a so-called "Pareto Front" in parameter space, and differ, essentially, on the weight that each solution gives to the different objectives of the optimisation problem. Our particular implementation of the evolutionary algorithm is tuned for diversity, so that nearly-equivalent or degenerate solutions will also appear composing the final population of solutions. In this way, the output of the algorithm gives a full picture of the optimisation problem, allowing the user to apply it as a guide to more in-depth analysis, such as selecting input layouts for a complete Monte Carlo simulation. 

In dealing with complex, hybrid arrays of Cherenkov Telescopes, the range of array configurations to be considered can be very large. Likewise, in specific cases when cost limitations or non-ideal site topology force a downgrade in the number of telescopes or a modification of the array from ideal solutions, such investigative tools might prove invaluable to assist the designer. 

Biological evolution interpreted as optimisation gives rise to a general theory in computation. Evolutionary Algorithms are based in the manipulation of a set of candidate solutions (individuals within a population) by random operations which mimic genetics. By describing the problem in terms analog to a genetic code, and operating on them with random "genetic operators", new generations of candidate solutions are produced and tested, following the principle of the "survival of the fittest". Fitness is measured in terms of the cost functions of the optimisation problem \cite{ref3.1}. 

In our implementation, the individuals to be evolved are groups of telescopes $t_i$ placed on a discrete grid, forming an array. Each telescope is described by its position $(x_i, y_i)$ in the grid and its type (for hybrid arrays). Each of these quantities are expressed in binary code, and the set of all these numbers for the $N$ telescopes that compose the array constitute the "genetic code" of the individual to be evolved. The algorithm can be adapted to treat the number of telescopes as an optimisation parameter as well.

The initial input population is a group of arrays randomly placed in the grid which will evolve to the final solution of optimal array layouts by means of random genetic operations on the binary digits (genes). The two basic operations are recombination -- the combination of the genetic material of two individuals (arrays) to form a new one -- and mutation : the alteration of the characteristics of an individual (array) by random deletion and/or replacement of some of its genes. Finally, the new individuals generated are selected by ranking their fitness and only the best solutions are propagated to the next generation. Thus, good genes are perpetuated, until the population converges, over several generations, to a stable set of optimised individuals. 

\begin{figure}
\includegraphics[width=\textwidth]{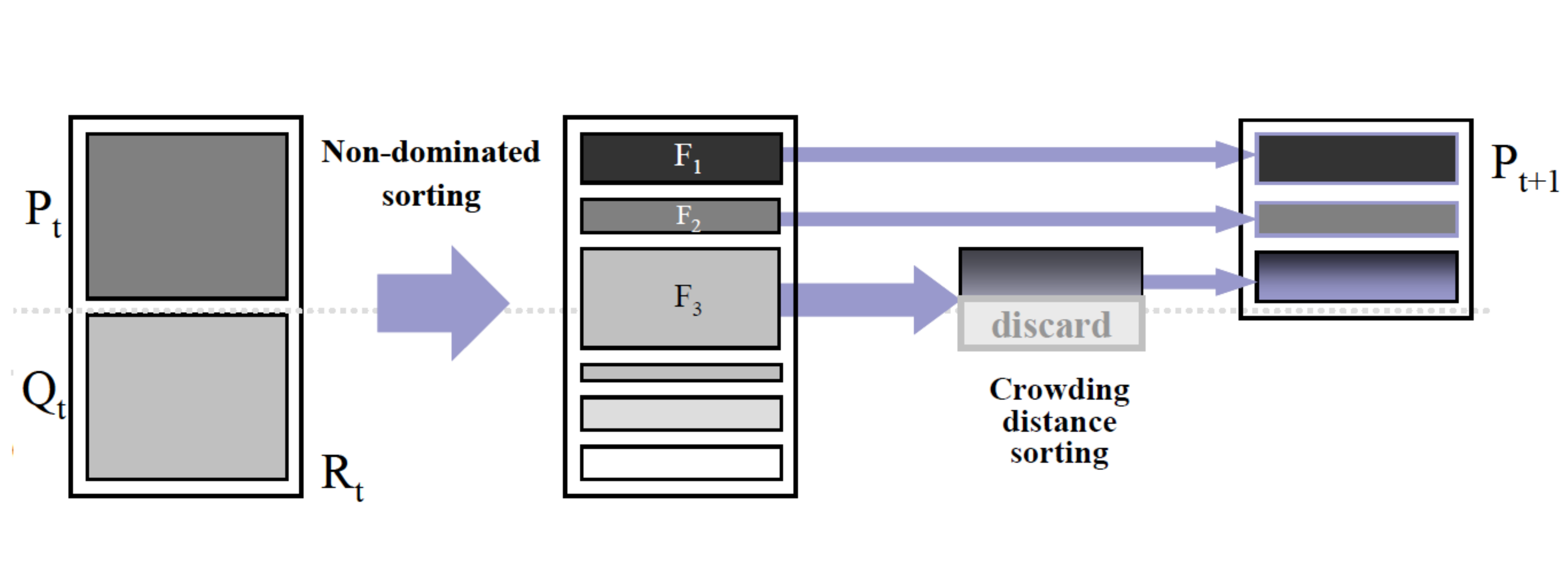}
\caption{Schematics of the evolutionary algorithm. Two initial populations $P_t$ and $Q_t$ are generated. The individuals compete and are ranked according to their fitness. The group of dominant solutions (better than any other solutions with respect to all metrics of the problem) survive directly to compose the next generation $P_{t+1}$. The remaining best, non-dominant solutions (better than other solutions in at least one metric) are sorted from, in order to guarantee maximum diversity: that is, they are selected to cover the broadest range of the parameter space. The remaining individuals are discarded, and the process is repeated in successive generations until the final population converges.}
\label{fig_ga}
\end{figure}

Figure~\ref{fig_ga} presents the schematics of our specific implementation of the algorithm. This particular version of the algorithm is designed to have fast convergence properties and good diversity. This means that the final solutions cover the largest possible extent of the optimum parameter space (a "Pareto Front", as will be illustrated in the following section), thus maximising the information returned from the algorithm.

\section{Preliminary Results and Conclusions}

\begin{figure}
\includegraphics[width=\textwidth]{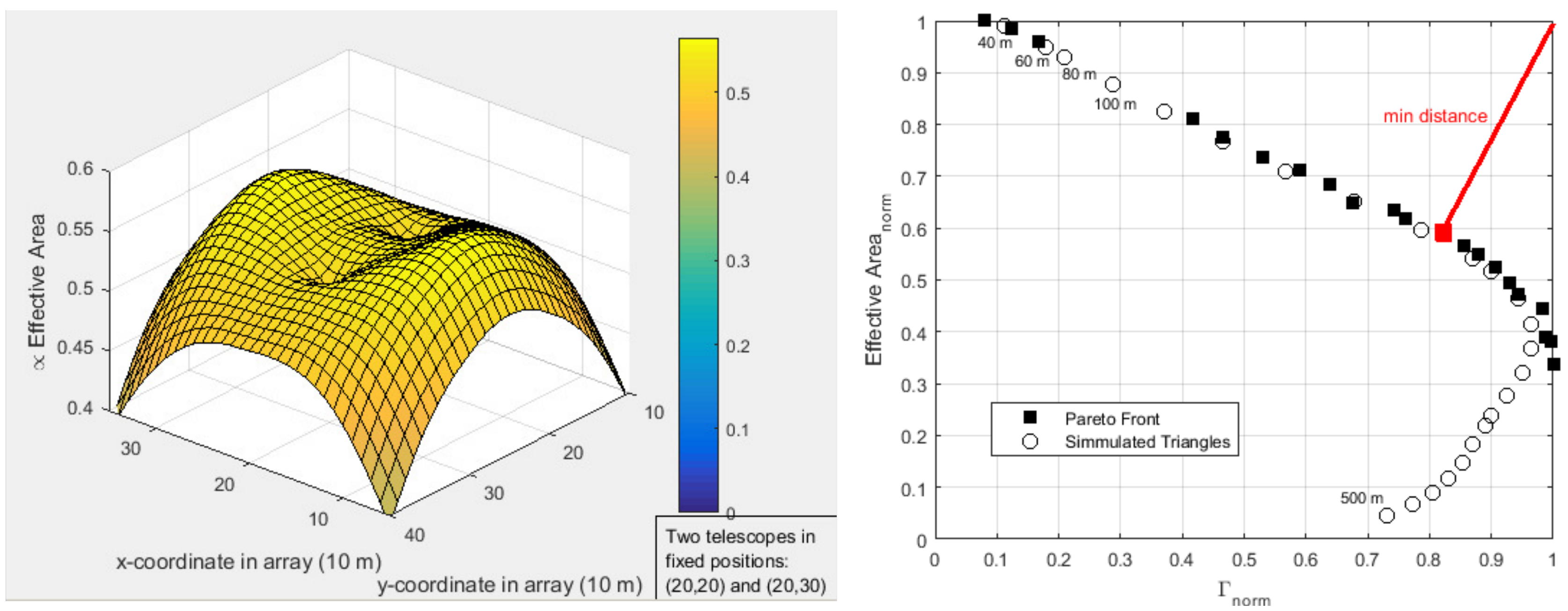}
\caption{(Left) Parametric space for the optimisation of the effective area of a 3-telescope stereoscopic array. The position of two telescopes are held fixed whereas the third telescope moves freely. The metric peaks at arrangements corresponding to near-equilateral triangles. (Right) The Pareto Front (black points) for the multi-parametric optimisation, showing the complete family of optimal solutions for the array layout. The point marked red is the best compromise solution, corresponding to the minimum distance, in normalised parametric space, to the maximum of the two metrics. The optimised solution is compared to simulated equilateral triangles of different sizes (white points).}
\label{fig_3telarray}
\end{figure}

In the following we show a few examples which illustrate the application of the Evolutionary Algorithm to the toy model array design. First of all, to illustrate how the evolutionary algorithm works, figure~\ref{fig_3telarray} (left) shows the parameter space for the effective area optimisation of an array of three telescopes, working in stereoscopic mode. Only the dependency on the metric (2.1) is presented, as two telescopes are held at fixed positions while the third telescope moves about. The metric peaks at arrangements corresponding to near-equilateral triangles with separation corresponding to the size of the Cherenkov light pool on the ground. It is important to observe that the genetic algorithm is not tuned to give exact, but near-optimal solutions according to the convergency criterion adopted. 

One of the strengths of the Evolutionary Algorithm is its capability to deal with multi-parametric optimisation simultaneously. This means that, as a result, it outputs a family of solutions, each corresponding to a different array layout, according to the weight given by each optimised objective. Figure~\ref{fig_3telarray} (right) shows the so-called Pareto Front for a simultaneous optimisation of a three-telescope array in the two metrics (2.1) and (2.2), presented in normalised scale. The point marked red, corresponding to the minimum distance, in normalised parametric space, to the maximum of the two metrics, is the best compromise solution. The optimised solutions of the Pareto Front are compared to simulated equilateral triangles of different sizes. Observe how, above a certain size of circa 250 m, the white points depart from the Pareto Front and its performance worsens in both objectives. Note also that the best compromise solution occurs at a value of $\Gamma$ corresponding to a nearly-equilateral triangle, with average stereoscopic angle $\sim$ 45-50$^\circ$.

The exact values where the metrics are maximised will depend on the models adopted for the Cherenkov lateral distribution and photon density. Here we adopt heuristic expressions for these, obeying the right functional forms and approximate scales, but not following any detailed Monte Carlos simulation. The sizes of the arrays and inter-telescope distance resulting from the optimisation should therefore be considered in an approximative way, and judged only for internal consistency, and not in comparison to other array design proposals.

\begin{figure}
\includegraphics[width=\textwidth]{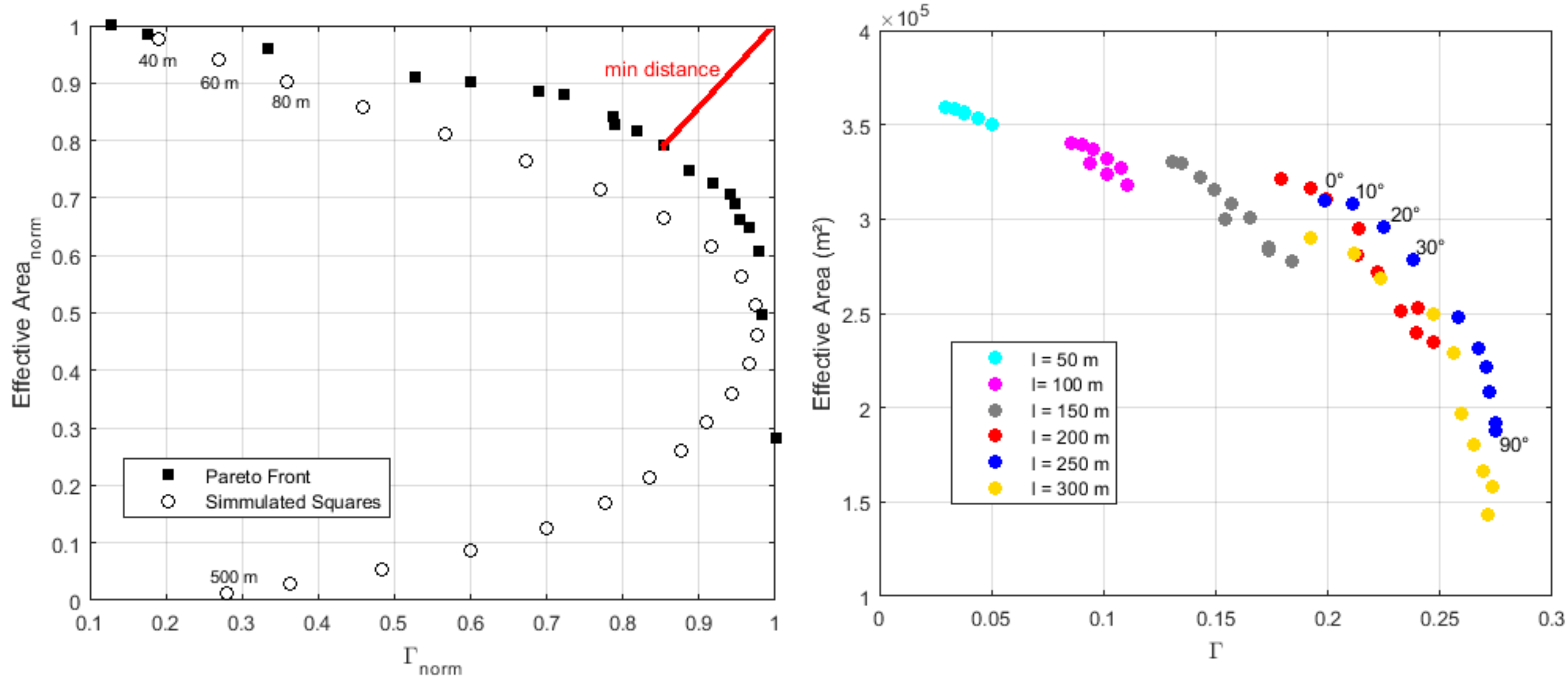}
\caption{(Left)The Pareto Front (black points) for the multi-parametric optimisation, showing the complete family of optimal solutions for the array layout of 4 telescopes, with minimum stereo multiplicity of $n=3$. The point marked red corresponds to the best compromise solution. The optimised family of solutions are compared to a set of simulated square configurations with different sizes. (Right) The plot shows the effect of changing each of the metrics (mean separation between pairs of instruments and average stereo angle between telescope pairs) to the position of the solution in parameter space.}
\label{fig_4telarray}
\end{figure}

As a second example, figure~\ref{fig_4telarray} shows a similar analysis for a 4-telescope array. Note how the Pareto Front is more distant from the simulated equilateral geometry (squares) than in the triangular case. The simulation indicates that, for the condition of minimum stereo multiplicity, $n=3$, losangos are the best geometric configuration. The different positions on the Pareto Front indicate different sizes and opening angles of the losango. The image on the right shows how variations on the size of the largest diagonal distance, $l$, and the losango opening angle, both affect the position of the solutions in the multi-parametric space. The optimal compromise solution is for an average distance between telescope pairs of 120 m and an average stereo angle between telescope pais of $\sim$ 55-60$^\circ$.

\begin{figure}
\includegraphics[width=\textwidth]{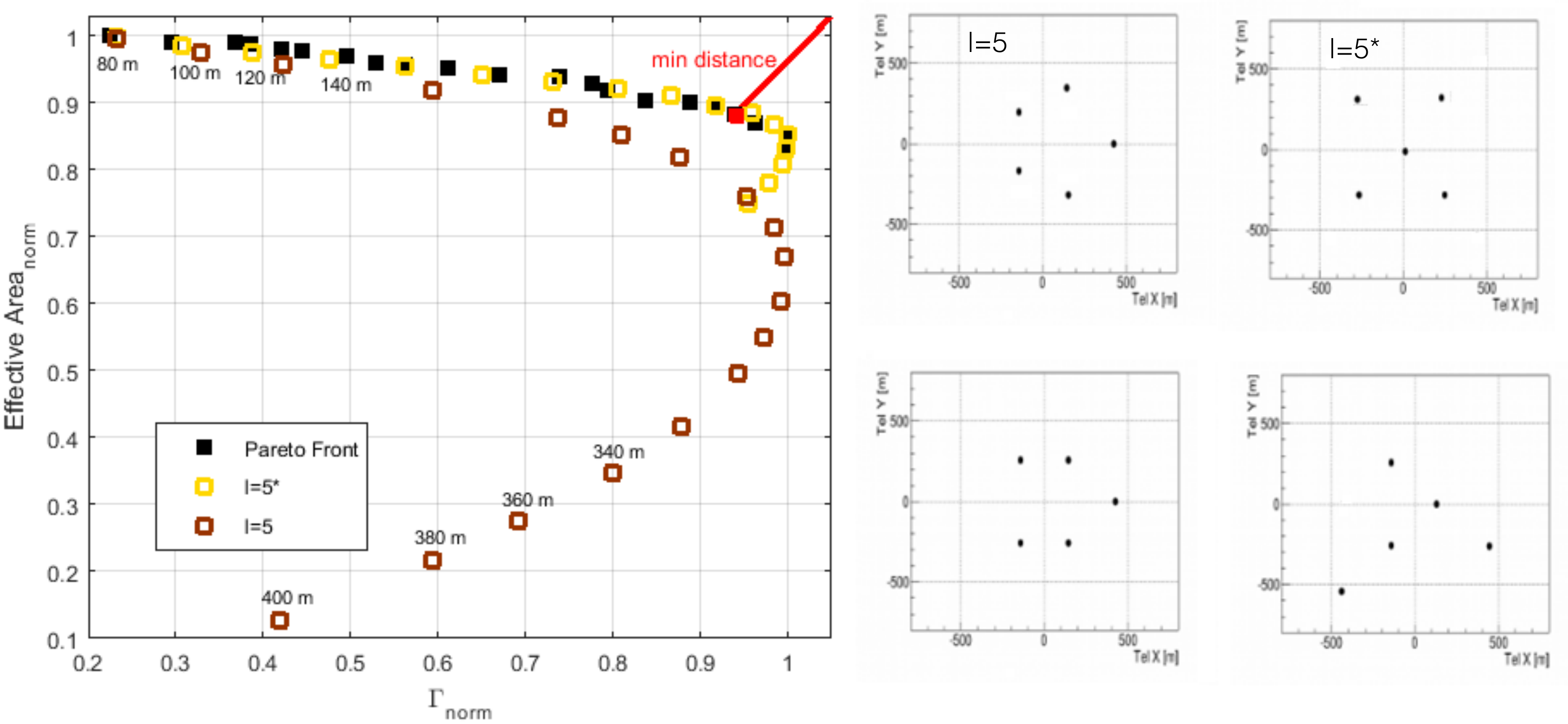}
\caption{(Left) The Pareto Front (black points) for the multi-parametric optimisation, showing the complete family of optimal solutions for an array layout of 5 telescopes. Observe how the Pareto Front falls within the locus of the $l=5^{*}$ solutions, corresponding to a configuration of 4-telescopes around a central fifth instrument, as opposed to the $l=5$ pentagonal solution. (Right) Examples of solutions given by the optimisation algorithm. Above are shown optimal pentagonal and $l=5^{*}$ configurations, whereas below non-optimal solutions explored but discarded by the algorithm are presented.}
\label{fig_5telarray}
\end{figure}

Finally, figure~\ref{fig_5telarray} shows the optimisation solution for another, less trivial case, of a 5 telescope array. Here the algorithm gives a strong indication of the best geometry to be investigated, favouring a configuration of 4-telescopes around a central fifth instrument over a regular pentagonal arrangement. This is clear from the left plot, where the Pareto Front is seen lying over the simulated $l=5^{*}$ set of figures, as opposed to the pentagons. 

This result can be readily understood in terms of the input model the algorithm is optimising: Since we are interested in multiplicity 3 stereo reconstructions, it is the configuration $l=5^{*}$ which gives the set of triangular sub-arrays with the largest average distance between telescope pairs (metric 2.1) and the largest average opening angle between telescope pairs (metric 2.2). 

The fact that we have analysed only simple cases here allows us to follow carefully what the algorithm is doing and how it is performing, reassuring ourselves of its capabilities and effective response to the optimisation problem. Nevertheless, as the physics introduced by new metrics becomes more complex or the number of telescope increases, or if we introduce hybrid arrays with different telescope types responding differently to each metric, an intuitive perception of how the geometry behaves in relation to the optimisation becomes less intuitive. This is where the algorithm becomes most useful as an exploratory guide for the array designer, investigating the parameter space of potential array configurations.

In conclusion, we have presented a working Evolutionary Algorithm applied to the problem of layout optimisation of an array of Cherenkov Telescopes. Although the algorithm is completely functional, work is ongoing in refining the metrics to improve the physics of the problem. More complex configurations, with additional optimisation parameters are being run to compose a future, more detailed publication and investigate arrays of an scale and complexity similar to that of CTA.

\end{document}